\documentclass[a4paper]{jpconf}
\usepackage{graphicx, soul,hyperref, slashed,amsmath,amsfonts,changepage,grffile,colortbl,ulem}
\usepackage{soul}
\usepackage[svgnames]{xcolor}

\DeclareRobustCommand{\eq}[1]{eq.~\eqref{eq:#1}}

\DeclareRobustCommand{\fig}[1]{Fig.~\ref{fig:#1}}

\DeclareRobustCommand{\sec}[1]{Sec.~\ref{sec:#1}}
\DeclareRobustCommand{\Sec}[1]{Section~\ref{sec:#1}}

\DeclareRobustCommand{\refcite}[1]{Ref.~\cite{#1}}

\newcommand{\cC}{\ensuremath{\mathcal{C}}}
\newcommand{\cH}{\ensuremath{\mathcal{H}}}
\newcommand{\cPT}{\ensuremath{\mathcal{PT}}}
\newcommand{\cP}{\ensuremath{\mathcal{P}}}
\newcommand{\cT}{\ensuremath{\mathcal{T}}}
\newcommand{\cK}{\ensuremath{\mathcal{K}}}
\newcommand{\cM}{\ensuremath{\mathcal{M}}}

\newcommand{\Veff}{\ensuremath{V_{\text{eff}}}}
\newcommand{\trace}{\ensuremath{\text{Tr}}}

\begin{document}

\begin{flushright}
MIT-CTP 5295
\end{flushright}
\begin{adjustwidth}{0pt}{-2cm}
\title{$\cPT\,$symmetry,$\,$pattern$\,\,$formation,$\,$and$\,\,$finite-density$\,$QCD}
\end{adjustwidth}

\author{Moses A. Schindler$^1$, Stella T. Schindler$^2$, and Michael C. Ogilvie$^1$}
\address{$^1$ Department of Physics, Washington University, St. Louis, MO 63130}
\address{$^2$ Center for Theoretical Physics, Massachusetts Institute of Technology, Cambridge, MA 02139}

\ead{stellas@mit.edu}

\begin{abstract}
A longstanding issue in the study of quantum chromodynamics (QCD) is its behavior at nonzero baryon density,
which has implications for many areas of physics.
The path integral has a complex integrand when the quark chemical potential is nonzero and therefore has a sign problem,
but it also has a generalized $\cPT$ symmetry.
We review some new approaches to $\cPT$-symmetric field theories, including both analytical techniques and methods for
lattice simulation. We show that $\cPT$-symmetric field theories with more than one field generally have a much richer phase structure than their Hermitian
counterparts, including stable phases with patterning behavior. The case of a $\cPT$-symmetric  extension of  a $\phi^4$ model
is explained in detail. The relevance of these results to finite density QCD is explained, and we show that a simple model of finite density
QCD exhibits a patterned phase in its critical region.

\end{abstract}

\section{Introduction}\label{sec:1}
Determining the phase diagram of quantum chromodynamics (QCD), the theory of the strong force, is an important goal with broad implications for nuclear and particle physics, astrophysics, and cosmology. The QCD phase diagram is often described as a function of the quark chemical potential $\mu$ and the temperature $T$. Decades of work have produced a sophisticated picture of the behavior of QCD at  nonzero temperature and zero chemical potential.
When $\mu \ne 0$, we are much less certain of the behavior of QCD. Lattice simulations with  $\mu\ne 0$ run into the sign problem: the Euclidean path integral has a complex integrand and does not have a probabilistic interpretation. Such sign problems may be non-deterministic non-polynomial (NP) hard in the general case \cite{Aarts:2015tyj}. Although much effort has been directed towards the development of algorithms to circumvent the sign problem  \cite{Aarts:2008rr,Cristoforetti:2012su}, we do not yet have a clear picture of QCD phase structure at finite density from lattice simulations. Astrophysical evidence relevant to QCD phase structure is equivocal at best, and exploration of the most important regions of the $\mu$-$T$ plane by direct experiments is still in its early stages.

\begin{figure}
\begin{center}
\includegraphics[width = 3 in]{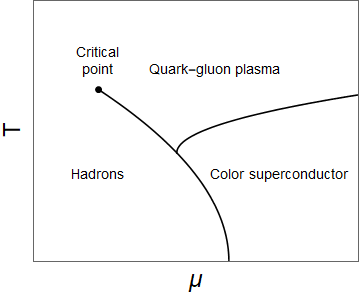}
\caption{A possible phase diagram of QCD in the $\mu$-$T$ plane. In the bottom left corner of the phase diagram, quarks and gluons are bound inside hadrons. In extreme conditions like compact stars and the early universe, it is hypothesized that quarks and gluons can become unbound and form exotic phases of matter. This paper focuses on the transition between hadrons and quark-gluon plasma, which is hypothesized to take the form of a first-order critical line that terminates in a second-order critical endpoint.}\label{fig:qcd-phase-diagram}
\end{center}
\end{figure}

Figure \ref{fig:qcd-phase-diagram} shows a possible phase diagram for QCD in the $\mu$-$T$ plane.
In the low-$T$, low-$\mu$ region, quarks and gluons are bound in hadrons. At very high $\mu$ and low $T$, there are strong arguments based on asymptotic freedom that QCD  must be a color superconductor \cite{Alford:2007xm}. However, we do not know where in the $\mu$-$T$ plane that behavior sets in. At high $T$ quarks and gluons are predicted to appear as a hot, dense plasma. There is good evidence from a variety of models that there is a first-order phase transition line emerging from the $T=0$ axis and terminating in a second-order critical point \cite{Fukushima:2010bq,Fu:2019hdw}. A number of experimental programs, including BES at RHIC \cite{Adamczyk:2017iwn} and CBM at FAIR \cite{Friman:2011zz}, are specifically targeting this critical behavior.

In this paper we look at finite-density QCD through the lens of non-Hermitian and $\cPT$ symmetric quantum theory. 
Although QCD at nonzero density and temperature has a path integral formulation with a complex integrand, the path integral is invariant under the combined action $\cC\cK$ of charge conjugation $\mathcal C$ and complex conjugation $\mathcal  K$, a form of $\cPT$ symmetry \cite{Meisinger:2012va}. 
We have developed new algorithmic and analytic techniques for studying $\cPT$-symmetric quantum field theories in order to better understand QCD. A key development has been the design of a new algorithm that allows for lattice simulation of $\cPT$-symmetric field theories \cite{Ogilvie:2018fov,Medina:2019fnx}. This algorithm recasts some models with complex actions into equivalent models with real actions. This has allowed us to compare lattice simulations with analytical results for models in the Z(2) symmetry class. In the process of this work, we have also learned about the relation of $\cPT$-symmetric field theories with complex actions to equivalent real representations, some of which are local and some of which are nonlocal. The latter are closely connected to the relationship between $\cPT$-symmetric and higher-derivative models.

The biggest surprise of our work has been the connection between $\cPT$-symmetric field theories and pattern formation. Pattern formation is known to occur in many systems at a wide range of length and energy scales in physics \cite{Seul476}. Patterns often arise from competition between attractive and repulsive forces.  Many examples are found in condensed matter physics; an exotic example is nuclear pasta, which has been predicted to arise in the inner crust of a neutron star. Here, competition between the Coulomb and nuclear forces leads protons and neutrons to cluster together in patterns which have been whimsically associated with pasta  (lasagna, spaghetti, gnocchi, etc.) \cite{Ravenhall:1983uh,10.1143/PTP.71.320, Caplan:2016uvu}.

$\cPT$-symmetric field theories provide natural realizations for some experimentally observed phenomena which are impossible in normal field theories.  Conventional field theories obey spectral positivity. For a Hermitian scalar field $\phi(x)$, this implies that its two-point function can be written as a sum over decaying exponentials
\begin{equation}
\langle\phi(x)\phi(y)\rangle=\int_0^\infty d\rho(m^2) \int {d^dk \over (2\pi)^d} {1\over k^2+m^2} e^{ik\cdot x}
\end{equation}
where $\rho(m^2)\ge 0$. This immediately rules out pattern formation, where order parameters exhibit periodic behavior. Spectral positivity also rules out propagators with sinusoidally modulated exponential decay. It is easy to see that such behavior is natural in $\cPT$-symmetric field theories when there are complex-conjugate eigenvalues of the Hamiltonian. As we will show in \sec{3}, there is a close relation between modulated exponential decay, pattern formation and second-order critical points in some $\cPT$-symmetric field theories.
Furthermore, there is a close connection between the balance of attractive and repulsive forces associated with pattern formation and the imaginary couplings often found in $\cPT$-symmetric models. Recall that in conventional field theories the force due to scalar particle exchange is always attractive; in contrast, an imaginary coupling in a $\cPT$-symmetric field theory leads to a repulsive coupling. In the model we study in \sec{3}, an imaginary coupling produces a Lifshitz instabilty which can be understood as the cause of pattern formation. This also strengthens the general connection between $\cPT$-symmetric models and higher-derivative theories.

In \sec{2}, we review some general results for $\cPT$-symmetric quantum field theories, including a new algorithm for simulating a large class of $\cPT$-symmetric scalar field theories. \Sec{3} explores the phase structure of $\cPT$-symmetric scalar field theories. A $\cPT$-symmetric extension of the conventional $\phi^4$ model is studied using both analytical methods and lattice simulation. \Sec{4} discusses the role of $\cPT$ symmetry in QCD at nonzero density. We summarize this work and offer concluding remarks in \sec{5}.

\section{$\cPT$ symmetry and quantum field theory }\label{sec:2}

In 1998, a groundbreaking paper by Bender and Boettcher showed that a complex extension of the harmonic oscillator, $\cH=-\partial_x^2 + x^2(ix)^\epsilon$, has a countably infinite spectrum of real and positive eigenvalues for $\epsilon \geq 0$ \cite{Bender:1998ke}. Bender and Boettcher observed 
that the invariance of this class of models under $\cPT$ symmetry, where $\cP$ is parity and $\cT$ is time reversal, is an important symmetry of $\cH$. We emphasize an important notational convention: in the context of non-Hermitian physics the letters $\cP$ and $\cT$ refer generically to any given pair of linear and antilinear operators; they need not represent parity inversion and time reversal operators specifically. 
Every eigenvalue of a $\cPT$-symmetric operator must be either real or part of a complex-conjugate eigenvalue pair. It has become conventional to say that an operator with entirely real eigenvalues has an unbroken $\cPT$ symmetry; otherwise the symmetry is said to be broken. This terminology is  statement about the reality of the full spectrum of excitations in a quantum-mechanical system, and is not directly related to spontaneous symmetry breaking in field theories. If one or more tuples of eigenvalues are coalesced at the same value, the operator is said to be at an exceptional point \cite{Bender:2007nj}. $\cPT$-unbroken Hamiltonians exhibit  many of the properties of Hermitian systems: unitarity, a positive norm under a ($\cC\cPT$) inner product \cite{Bender:2007nj}, eigenfunctions that possess analogs of interlacing zeros \cite{Bender:2000wj,Schindler:2017suy}, orthogonality, and completeness \cite{Mezincescu:2000fd,Weigert:2003py}. These behaviors break down in a characteristic manner when moving a parameter like $\epsilon$ in $\cH$ past an exceptional point and into the broken symmetry regime. Non-Hermitian systems without $\cPT$ symmetry do not in general exhibit any of these properties. For further details, see \refcite{Bender:2019cwm}.

The first experimental realizations of systems with $\cPT$ symmetry were in optical waveguides with gain and loss \cite{El-Ganainy:07, PhysRevLett.103.093902,ruter2010}. The paraxial wave equation of optics is isomorphic to the Schr{\"o}dinger equation, albeit with the electric field $E$ taking the place of the quantum-mechanical wavefunction $\psi$, refractive index $n(x)$ instead of potential $V(x)$, and propagation direction $z$ replacing time $t$. We can understand the appearance of an imaginary, or non-Hermitian, term in a Hamiltonian as representing the exchange of energy between a system and its environment, with the direction of energy flow indicated by the term's sign. In the $\cPT$-symmetric Schr{\"o}dinger equation, the potential term must satisfy Im $V(x) = -$ Im $V^*(-x)$, which means the system loses precisely as much energy to its environment as it gains back. Researchers have studied $\cPT$ symmetry in many areas of physics; see, e.g., the reviews \cite{feng2017,Christodoulides2018,Miri2019}. 

$\cPT$-symmetric quantum field theories can have many analogous properties to conventional quantum field theories \cite{Bender:2019cwm}. For example, the spectrum of a $\cPT$-QFT may be real and bounded below. It is possible to construct an inner product under which a $\cPT$-quantum field theory exhibits unitary time evolution \cite{PhysRevLett.93.251601}. This has implications for important physics models; for example, when one analyzes the Lee Model in a $\cPT$-symmetric framework, there is no ghost; under a correctly defined inner product, all states have positive norm \cite{Bender:2004sv}. 



The path integral approach to quantum mechanics and quantum field theory is a powerful tool, but the development of $\cPT$ symmetric quantum theories has largely focused on the canonical approach emphasizing Hilbert space structure. We have developed a Euclidean path integral technique which recasts a broad class of $\cPT$-symmetric scalar field theories with complex actions into real forms. If this real form of the functional integral
has a positive integrand, then the sign problem is removed and the path integral can be simulated with standard lattice methods.

As an example, let us consider the general case of a single $\cPT$-symmetric scalar field $\chi(x)$ with action
\begin{equation}
S(\chi)=\sum_{x}\left[\frac{1}{2}(\partial_{\mu}\chi(x))^{2}+V(\chi(x))-ih(x)\chi(x)\right],
\end{equation}
where $V(\chi)=V^*(-\chi)$. 
In the lattice field theory, we take the spatial derivative term to be the standard finite-difference expression.
The external
field $h(x)$ is dependent on Euclidean spacetime location, and is used to deduce correlation functions
for $\chi$ in the new representation. 
We now rewrite the kinetic and potential terms as Fourier transforms.
We write the single-site kinetic term as:
\begin{equation}
\exp\left[\frac{1}{2}\left(\partial_\mu\chi\right)^{2}\right]=\int d\pi_{\mu}(x)\exp\left[\frac{1}{2}\pi_{\mu}(x)^{2}+i\pi_{\mu}(x)\partial_{\mu}\chi_{x}\right].
\end{equation}
The weight term $w[\chi(x)]\equiv\exp\left[-V(\chi(x))\right]$
associated with the potential $V$ can be written as the transform
of a real dual weight $\tilde{w}[\tilde{\chi}(x)]$.
If $\tilde{w}>0$, then we can define the real dual potential $\tilde{V}(\tilde{\chi})\equiv-\log[\tilde{w}(\tilde{\chi})]$.
This condition of dual weight positivity is equivalent by Bochner's theorem to the condition that the weight $w[\chi(x)]$
is positive definite.
After a lattice integration by parts, it is straightforward to integrate
out $\chi$ to obtain the action
\begin{equation}
Z=\int\prod_{x}d\pi_{\mu}(x)\exp\left\{ -\sum_{x}\left[\frac{1}{2}\pi_{\mu}^{2}(x)+\tilde{V}(\partial\cdot\pi(x)-h(x))\right]\right\},
\end{equation}
which is positive everywhere. In such a representation, standard lattice
simulation methods may be used. This representation is local and may
be used in any dimension; algorithmic implementation is easy. The
extension to multiple scalar fields is trivial.
This transformation can be understood as a kind of duality transform appropriate for real-valued
fields.
The dual positive weight condition puts restrictions on potentials $V$ that can be simulated using this method.
However, $\tilde V$ need only be real for $V$ to be $\cPT$ symmetric, so this method may
be used to create an infinite number of $\cPT$-symmetric field theories 
in any space-time dimensionality.

This technique is quite similar to the method used in the analysis of the wrong-sign $x^4$ model, where the functional
integration contour is complex, lying in a nontrivial Stokes wedge \cite{Bender:2006wt}.
The method we have presented applies to scalar field theories when the functional integration contour is along the real axis.
It seems likely that similar techniques can be used to reduce some models defined on complex contours to a real form.
\clearpage
\section{Phase structure in $\cPT$-symmetric field theories}\label{sec:3}

In this section we review recently developed techniques to determine the phase diagram of a $\cPT$-symmetric
scalar quantum field theory and the application of those techniques to a
 $\cPT$-symmetric extension of Hermitian $\phi^4$ theory, providing some additional details on the phase structure
 \cite{Medina:2019fnx}.
For standard scalar field theories, there is a well-defined procedure for studying phase structure within the path integral formalism using perturbation theory. Writing the Lagrangian density $\mathcal L$ in terms of a set of real fields $\phi^a(x)$ where $a$ runs from $1$ to $N$, we have:
\begin{equation}\label{eq:Hermitian-lagrangian}
\mathcal{L}\left(\phi^a\right) = \frac{1}{2}\left(\partial_\mu\phi^a\right)^2  + V(\phi^a).
\end{equation}
If the potential $V$ is bounded from below, there is always at least one spacetime-independent solution $\phi^a_0(x)$ which achieves the global minimum of the potential, and is thus a candidate for building a perturbative solution around. The mass matrix is real and symmetric, and has non-negative eigenvalues, with zero eigenvalues associated with Goldstone bosons.

This can change profoundly in a $\cPT$-symmetric scalar field theory with a complex action. Consider such a theory in $d$ dimensions.
We suppose there is a set of fields $\phi^a$ that transform trivially under $\mathcal P$ and $\mathcal T$
transformations and a set of fields $\chi^b$ that transform nontrivially, such that the action
is invariant under the combined action of the operators $\cPT$.
We find a perturbative solution $(\phi^a_0,\chi^b_0)$  by minimizing $V$
over the class of spacetime-independent solutions.
We assume that $\mathcal{PT}$ symmetry is maintained, which implies that $\phi^a_0$ is real,
$\chi^b_0$ is imaginary and $V(\phi_0,\chi_0)$ is real.
The mass matrix $\mathcal{M}$ associated with this solution is given in block form by 
\begin{equation}
{\arraycolsep=1.4pt\def\arraystretch{2.4}
\mathcal{M}=\left(\begin{array}{cc}
\cfrac{\partial^{2}V}{(\partial\phi^{a})^2} & \cfrac{\partial^{2}V}{\partial\phi^a\partial\chi^b}\\
\cfrac{\partial^{2}V}{\partial\phi^a\partial\chi^b} & \cfrac{\partial^{2}V}{(\partial\chi^b)^{2}}
\end{array}\right)}
\end{equation}
evaluated at $(\phi^a_0,\chi^b_0)$.
This mass matrix is not necessarily 
Hermitian but is $\mathcal{PT}$-symmetric:
\begin{equation}
\mathcal{M} = \Sigma \mathcal{M}^* \Sigma,
\end{equation}
where $\Sigma$ is a diagonal matrix with entries $+1$ associated with the $\phi^a$ fields and $-1$ associated with
the $\chi^b$ fields.
The characteristic equations for $\mathcal M$ and $\mathcal M^*$ are the same, so they have the same eigenvalues. 
As a consequence, every eigenvalue of $\mathcal{M}$ must be either real or the complex-conjugate of another eigenvalue.

The zeros of $\det \left( q^2 + \mathcal{M} \right)$ are the poles in momentum space of the matrix propagator
for the $\phi$ and $\chi$ fields. For standard quantum field theories,  stable perturbative vacua have propagator poles 
that are real and negative as a function of $q^2$, so $q^2>0$ implies $\det \left( q^2 + \mathcal{M} \right)>0$ in this case.
Poles with $q^2<0$ lead to exponentially decaying propagators, consistent with spectral positivity. In $\cPT$-symmetric
models, complex-conjugate poles lead to sinusoidally-modulated exponential decay, which is inconsistent with spectral positivity.
This behavior is familiar from $\cPT$ symmetry breaking in quantum mechanics, where by definition energy eigenvalues become complex.
Such behavior has been known for some time in condensed matter physics, where the boundary between two parameter regions where the propagators change behaviors is known as a disorder line \cite{PhysRevB.1.4405}. We use the term disorder line to avoid confusion in models that have first-order critical lines and second-order critical endpoints.

The one-loop effective potential of our generic $\cPT$-symmetric scalar field theory at a classical homogeneous solution
is given by
\begin{equation}
\Veff(\phi_0,\chi_0)=V(\phi_0,\chi_0)+{1 \over 2}\int {d^d q \over (2\pi)^d} \log \det \left( q^2 + \mathcal{M} \right).
\end{equation}
If $\det \left( q^2 + \mathcal{M} \right)$ becomes negative for real $q^2$, the one-loop term develops an imaginary part,
indicating an instability of that particular solution.
If the eigenvalues of the mass matrix are all either positive or in a complex-conjugate pair, we have $\det \left( q^2 + \mathcal{M} \right)>0$,
and the classical solution $(\phi_0,\chi_0)$ is stable against small fluctuations. Such solutions may be stable or metastable, but they are not unstable.
If a homogeneous solution has a mass matrix with negative roots, then $\Veff(\phi_0,\chi_0)$ will develop an imaginary part,
indicating instability.
The most interesting case occurs when $\det\,\cM>0$ but det$\,(q^2+\cM)$ has an even number of positive zeros. 
The condition $\det\,\cM>0$ tells us that the solution is stable against $q^2=0$ fluctuations, but 
the propagator is unstable to fluctuations for some values of $q^2>0$. When a homogeneous solution is not stable, we might consequently assume the full theory is unstable. But there is a second possibility: simply, that the stable solution is not homogeneous. This is what happens when there are an even number of positive zeros: the ground state field configurations are patterned. In normal field theories, such phases are not associated with stable vacua because they do not represent global minima of the vacuum energy density. In $\cPT$-symmetric theories, a homogeneous solution may be the global minimum across all homogeneous solutions but be unstable against $q^2 \ne 0$ fluctuations.  We summarize our conjectured phase structure for $\cPT$-symmetric field theories in Table 1.

\setlength{\tabcolsep}{0.5em}
{\renewcommand{\arraystretch}{1.3}
\begin{table}
\begin{center}
\begin{tabular}{|p{0.75 in}p{0.6 in}p{1.75 in}p{2.3 in}|}
\hline
Region & det($\cM$)&Zeros of det($q^2+\cM$)&Behavior of the propagator\\
\hline
\rowcolor{AntiqueWhite} Normal & Positive & All zeros are negative & Exponential decay\\ 
\rowcolor{LightCyan}$\cPT$ broken & Positive & One or more pairs of zeros are complex conjugates & Modulated exponential decay\\
\rowcolor{Honeydew}Patterned & Positive & Even \# of positive zeros & Stable to $q^2=0$ fluctuations, but unstable to $q^2>0$ fluctuations\\
\rowcolor{MistyRose} Unstable & Positive&Odd \# of positive zeros&Instability to $q^2=0$ fluctuations\\
\rowcolor{MistyRose} Unstable & Negative&---&Instability to $q^2=0$ fluctuations \\
\hline
\end{tabular}
\end{center}
\caption{\label{tbl:propagator} The phase diagram of a non-Hermitian $\cPT$-symmetric quantum field theory, described in terms of the behavior of its matrix propagator. Note that two new behaviors can occur in $\cPT$-QFTs that are generally not present in conventional Hermitian QFTs. First, the propagator may possess complex zeros, and due to $\cPT$ symmetry these zeros must come in complex-conjugate pairs. This leads to a propagator with exponential decay that is modulated by a sinusoidal term, and the propagator may even be partly negative. Nonetheless this complex region exhibits stable behavior and field configurations with a typical appearance. There is also a region with tachyonic modes that can support stable ground-state field configurations, although these field configurations are not homogeneous.}
\end{table}
}

\subsection{A $\cPT$-symmetric model with Z(2) symmetry}\label{sec:3-1}

We now consider a $\cPT$-symmetric extension of a conventional $\phi^4$ model which is amenable to both analysis and lattice simulation: 
\begin{equation}\label{eq:action}
S(\phi,\chi) = \sum_x \frac{1}{2}(\nabla_\mu \phi)^2 + \frac{1}{2}(\nabla_\mu \chi)^2+ V(\phi,\chi),
\end{equation}
where we set 
\begin{equation}
V(\phi,\chi) =  \frac{1}{2}m_\chi^2 \chi^2   -ig\phi\chi+ U(\phi)+h\phi.
\end{equation} 
The potential $U$ for $\phi$ is the double-well potential $U(\phi)=\lambda(\phi^{2}-v^{2})^{2}$.
Equation (\ref{eq:action}) represents a Hermitian scalar field $\phi(x)$ coupled to a $\cPT$-symmetric scalar field $\chi(x)$ by the imaginary strength $ig$.
The model is invariant not only under the Z(2) symmetry $(\phi, \chi)\rightarrow (-\phi, -\chi)$, but also under the generalized $\cPT$ symmetry of complex conjugation
combined with $(\phi, \chi)\rightarrow (\phi, -\chi)$.
$\cPT$-symmetric models similar to this one can easily be constructed for other universality classes besides Z(2).
For example, an O($N$)-symmetric model with an $N$-dimensional Hermitian field $\vec \phi$ can be extended
to a $\cPT$-symmetric model O($N$) model with an additional field $\vec \chi$ and an imaginary term in the Lagrangian
of the form $-ig\vec\phi\cdot\vec\chi$.

Because $\chi$ enters quadratically in the action $S$, it can easily be integrated out of the functional integral, giving
a nonlocal effective action of the form
\begin{equation}
S_{\text{eff}}=\sum_{x} \left[\frac{1}{2}(\partial_{\mu}\phi(x))^{2}+\lambda(\phi^{2}-v^{2})^{2}+h\phi\right]\nonumber
+\frac{g^{2}}{2}\sum_{x,y}\phi(x)\Delta(x-y)\phi(y),
\end{equation}
where
\begin{equation}
\Delta(x-y)=\int {d^dq \over (2\pi)^d} {1 \over q^2 + m_\chi^2} \,e^{i q \cdot (x-y)}.
\end{equation}
This model has been extensively studied in the case $m_\chi = 0$, where it is known to give rise
to pattern-forming regions;
see, {\it e.g.}, \cite{PhysRevE.66.066108} and references therein.
The $m_\chi = 0$ limit is sometimes described in the condensed matter literature as Coulomb frustrated because the extra interaction acts against the symmetry-breaking behavior of the  $\phi^4$ model 
\cite{PhysRevE.66.066108,PhysRevLett.100.246402,ORTIX2009499}.

The value of the order parameter at tree level,  $\phi_0$, is determined by minimizing the potential $V$
or equivalently by minimizing the effective potential associated with $S_{\text {eff}}$: 
\begin{equation}
V_{\text{eff}}(\phi_0)=\lambda(\phi_0^2-v^2)^2+g^2\phi_0^2/2 m_\chi^2-h\phi_0. \label{eq:v-eff}
\end{equation}
The effect of $\chi$ on $\phi_0$ for $h=0$ is to restore the symmetric value $\phi_0 = 0$  at sufficiently large
values of $g$. However,  $\phi_0$ tells us the value of the zero-momentum component of $\phi$.
Pattern formation is associated with Fourier modes $\tilde\phi\left(q\right)$ with nonzero $q$, which requires additional calculation.
One approach to understanding pattern formation is to expand $S_{\text {eff}}$ in a derivative expansion. The last, nonlocal term in $S_{\text{eff}}$ generates an infinite series of local higher-order derivative terms
\begin{equation}
\int d^4 x\, {g^2 \over 2 m_\chi^2}\phi(x)\sum_{n=0}^{\infty}  \left( {-\nabla^2 \over m_\chi^2 }\right)^n 
\phi(x).  
\end{equation}
The $n=0$ term in this expansion is the last term in $V_{\text {eff}}$. The $n=1$ represents a correction to the kinetic term, and higher $n$ terms give
higher-derivative terms in $S_{\text {eff}}$.
Crucially, the $n=1$ term of the series is negative, indicating that the quadratic derivative term of $S_\text{eff}$ becomes negative for sufficiently large $g$. When the quadratic kinetic term is sufficiently negative, the homogeneous phase is unstable to perturbations with nonzero wave number.  Thus the occurrence of a pattern-forming region is a manifestation of a Lifshitz instability \cite{cha95}.

A global picture of the phase structure follows from the inverse $\phi$ propagator obtained from $S_{\text {eff}}$ at tree level:
\begin{equation}
G^{-1}(q^2)=q^2 +U''(\phi_0)+{{g^2} \over{q^2+m_\chi^2}}
\end{equation}
where $U''(\phi_0)=-4\lambda v^2 +12\lambda \phi_0^2$.
The allowed phases of the model are determined using $\phi_0$ and $G(q)$.
The poles of the propagator $G$ are zeros of
$(q^2+m^2_\chi)(q^2+U)+g^2=0$.
This quadratic in $q^2$ has real coefficients, so its roots are either both real or form a complex-conjugate pair.
Specifically, we can write the poles of $G(q^2)$ as $q^2 = r_\pm$ with
\begin{equation}\label{eq:example-roots}
r_\pm = -\frac{1}{2}\left(m_\chi^2  -4\lambda v^2 + 12\lambda \phi_0^2\right)\pm \frac{1}{2} \sqrt{(m_\chi^2 + 4\lambda v^2 - 12\lambda \phi_0^2)^2 - 4g^2}.
\end{equation}
If $r_-<0$ and $r_+>0$, the position-space propagator grows exponentially and the system is unstable. If $r_+, r_-<0$ then the position-space propagator decays exponentially and our system behaves like a normal, stable quantum field theory. If $r_+ = r_-^*$ then the position-space propagator decays exponentially with sinusoidal modulation but the homogeneous vacuum is also stable in this region of the phase diagram. The condition $r_+,r_->0$ leads to pattern formation. 

An equivalent approach to determining the phase structure is to start from $S$ rather than $S_{\text{eff}}$ and find the static solution
$\left(\phi_0,\chi_0\right)$  which minimizes $V(\phi,\chi)$. Unless the underlying $\cPT$ symmetry of $S$ is broken,
$\phi_0$ will be real and $\chi_0$ will be purely imaginary. Linearizing the propagator around the static solution, we find the inverse propagator for the $\left(\phi,\chi \right)$ set of fields is $q^2+\mathcal{M}$, where $\mathcal{M}$ is the $2\times 2$ mass matrix
\begin{equation}
{\arraycolsep=1.4pt\def\arraystretch{2.4}
\mathcal{M}=\left(\begin{array}{cc}
\cfrac{\partial^{2}V}{\partial\phi^{2}} & \cfrac{\partial^{2}V}{\partial\phi\partial\chi}\\
\cfrac{\partial^{2}V}{\partial\phi\partial\chi} & \cfrac{\partial^{2}V}{\partial\chi^{2}}
\end{array}\right),}
\end{equation}
which is
\begin{equation}
{\mathcal M}=\left(\begin{array}{cc}
U''(\phi_0) & ig\\
ig & m_{\chi}^{2}\\
\end{array}\right).
\end{equation}
The mass matrix $\mathcal M$ is not Hermitian, but it satisfies a $\cPT$ symmetry condition 
\begin{equation}
M = \sigma_3 M^* \sigma_3.
\end{equation}
As in our general analysis, this condition implies that the eigenvalues of $\mathcal M^*$ must be the same as those of $\mathcal M$, and thus they are either both real or form a complex pair. The zeros of the inverse matrix propagator can be obtained as the zeros of the characteristic equation
\begin{equation}
\det(q^2 + \mathcal{M}) = (q^2)^2+q^2\, \trace(\mathcal{M})  + \det(\mathcal{M})=0
\end{equation}
The coefficients $\trace(\mathcal{M})$ and $\det(\mathcal{M})$ are real, implying that the roots are either both real or form a complex-conjugate pair. The zeros of the characteristic equation are the propagator poles and so the two methods give the same results.

\begin{figure}
\begin{center}
\includegraphics[width = 3 in]{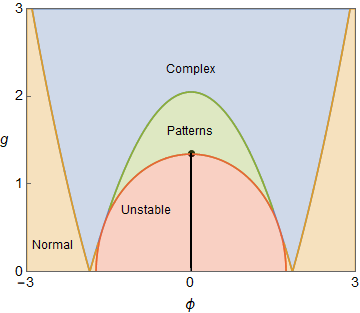}
\includegraphics[width = 3 in]{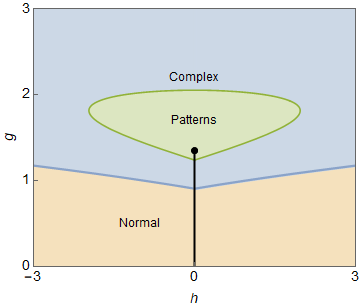}
\caption{The phase diagram of \eq{action}, plotted in two different parameter spaces.
In the $\langle \phi \rangle-g$ plane, the critical line runs up the $g$ axis until it terminates in a critical endpoint where it intersects the disorder line that marks the boundary between the patterned and the complex regions. 
When we make a change of variables to the $g-h$ plane there is no unstable region, and the critical point lies within the patterned region.
}\label{fig:phases2ways}
\end{center}
\end{figure}

In Figure \ref{fig:phases2ways}, we plot the regions for the four distinct phases of the model as a function of
$\langle \phi \rangle$ for the parameter set $m^2=1/2$, $\lambda=1/10$ and $v=3$. The graph on the left shows the phase diagram in the $\phi$-$g$ plane, while the graph on the right shows the phase structure in the $h$-$g$ plane. The different regions are classified by the nature of the poles of the $\phi$ propagator in the $q^2$ complex plane. We denote the region of parameter space where both poles are real and negative as ``Normal'' (orange in \fig{phases2ways}). This leads to exponential decay of the $\phi$ propagator, as it does in conventional field theories. In the region labeled Complex (blue), the poles as a function of  $q^2$ in the $\phi$ propagator are complex conjugates. This region is similar to the so-called broken $\cPT$ region of $\cPT$-symmetric quantum mechanical models. The $\phi$ propagator in this region also decays exponentially, but with sinusoidal modulation. This behavior violates spectral positivity and does not occur in Hermitian models. This behavior in $\cPT$-symmetric field theories is the field theory analog of complex-conjugate energy eigenvalues in $\cPT$-symmetric quantum mechanics. The boundary between the Normal and Complex regions is called a disorder line. The region labeled Patterns (green) is the region where both poles are real and positive; it is in this region where persistent patterns occur.  In the Unstable region (red), both poles are real with one positive and one negative. This region is not thermodynamically stable. It is inaccessible as an equilibrium state in the canonical ensemble where $h$ is a free parameter.

The phase diagram in the $h$-$g$ plane shows a cut at $h=0$, as in the familiar case of a ferromagnetic Ising model. This is a first-order phase transition, across which $\phi_0$ jumps. The two sides of the cut may be in the Normal, Complex or Pattern regions, as shown in the figure. The unstable phase does not appear in the $h$-$g$ phase diagram, and there are also metastable states from the other three phases which do not appear. These unstable and metastable states have moved through the cut into an analytic continuation from the stable regions. The critical line terminates at a second-order critical point where the $\phi$ field becomes massless. In the $\phi$-$g$ plane, this occurs at the boundary between the unstable and patterned regions, but in the $h$-$g$ plane, the metastable pattterned states are lost, and the critical end point appears in the middle of the patterning phase.

\begin{figure}
\begin{center}
\includegraphics[width = 3 in]{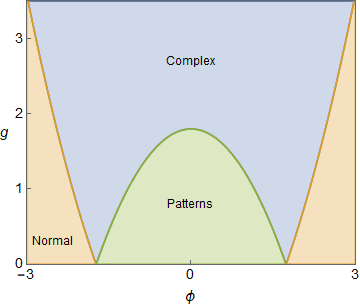}
\caption{The phase diagram of \eq{action} in the $m_\chi^2=0$ limit, where all other parameters 
are the same as in Figure \ref{fig:phases2ways}. Note that the unstable region has completely
disappered for $g>0$ in this limit.
}\label{fig:m2eq0}
\end{center}
\end{figure}

The location of the boundary between the patterned and unstable regions in the $\phi$-$g$ plane is controlled by the parameter
$m_\chi^2$. If we take $m_\chi^2\rightarrow \infty$, then the effects of the $\chi$ field on $\phi$ disappear, and both the complex and patterned phases disappear. In this limit, the phase diagram is that of the usual $\phi^4$ model. On the other hand, the limit $m_\chi^2\rightarrow 0$ turns the Yukawa (screened Coulomb) potential induced by $\chi$ into the long-range Coulomb potential. This is a smooth limit in the $\phi$-$g$ plane and in the phase diagram in \fig{m2eq0}. The unstable region disappears in this limit. Similarly, the pattern-forming region disappears in the limit $m_\chi^2\rightarrow \infty$.

\begin{figure}
\begin{center}
\includegraphics[width = 3 in]{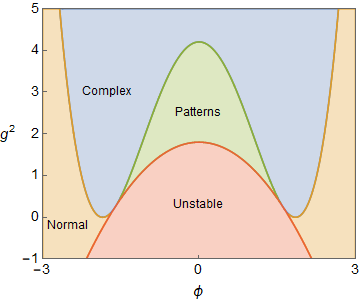}
\caption{The phase diagram of \eq{action} in the $\phi$-$g^2$ plane, where all parameters are the same as in Figure \ref{fig:phases2ways},
but the phase diagram has been extended to $g^2<0$. For $g^2<0$ there are only normal and unstable phases, as expected for a Hermitian theory. Note that the small region of normal behavior sandwiched between the complex and unstable regions in \fig{phases2ways} is connected to the larger normal region, when viewing the system in the extended $g^2$ plane.
}\label{fig:g2phase}
\end{center}
\end{figure}

The effective action $S_{\text{eff}}$ is a function of $g^2$, and can be continued to $g^2<0$. This corresponds to the continuation $g \rightarrow ig$ in $S$, in which case the action $S$ becomes real, and neither pattern formation nor complex $q^2$ poles can occur. The small areas of normal behavior seen in \fig{phases2ways} in between the complex and unstable phases are connected to the larger normal region when the phase diagram is plotted as a function of $g^2$ and $g^2<0$ values are included as in \fig{g2phase}. In the limit $m_\chi^2\rightarrow 0$, the unstable region occurs only for $g^2<0$.


The behavior we have found is based on the analysis of  $S_{\text{eff}}$  at tree-level, and is thus independent of dimensionality. As we know, 
fluctuations at one loop and beyond can strongly effect critical behavior in a dimension-dependent way.
We know that in the simple $\phi^4$ model, fluctuations destroy spontaneous symmetry breaking below $d=2$, the lower critical dimension for the model. Examination of the infrared behavior of the $\phi$ propagator
in our model shows that it generally has infrared behavior no worse than  $1/q^2$, as in the simple $\phi^4$ model. The only exception
is the point where both masses are zero, which occurs at tree level when $\trace (\mathcal {M})=0$ and   $\det(\mathcal{M})=0$. This
does not change the overall phase diagram for $d \ge 2$. A similar analysis applies in other models.


\subsection{Simulations of the Z(2)-symmetric model}\label{sec:3-2}

We can complement our analysis with lattice simulations. Using the method described in \sec{2}, we cast the action for $\cPT$-extended $\phi^4$ theory in \eq{action} into an entirely real, positive form
\begin{equation}\label{eq:simulatable-example}
\tilde{S} = \sum_x \left[\frac{1}{2} (\nabla_\mu\phi)^2 + \frac{1}{2}\pi_\mu^2 + \frac{1}{2m_\chi^2}(\nabla\cdot \pi - g\phi)^2 + \lambda(\phi^2-v^2)^2 + h\phi\right],
\end{equation}
which can be simulated using the Metropolis algorithm.

We have performed an extensive set of simulations of the model in two dimensions using the Metropolis algorithm applied to the real action $\tilde S$. The parameters $m_\chi^2$, $\lambda$, and $v$ were fixed to the same values shown in the phase diagrams of the previous section:  $m_\chi^2 = 0.5$, $\lambda = 0.1$ and $v = 3$. The simulations shown here were performed using a $64\times 64$ lattice, typically with a hot start followed by 20,000 sweeps after a period of equilibration. Pattern formation has also been observed in three dimensions \cite{Ogilvie:2018fov,Medina:2019fnx}, but the two-dimensional case is much easier to visualize, and our analytical results are independent of dimension.

\begin{figure}
\begin{center}
\includegraphics[width = 6 in]{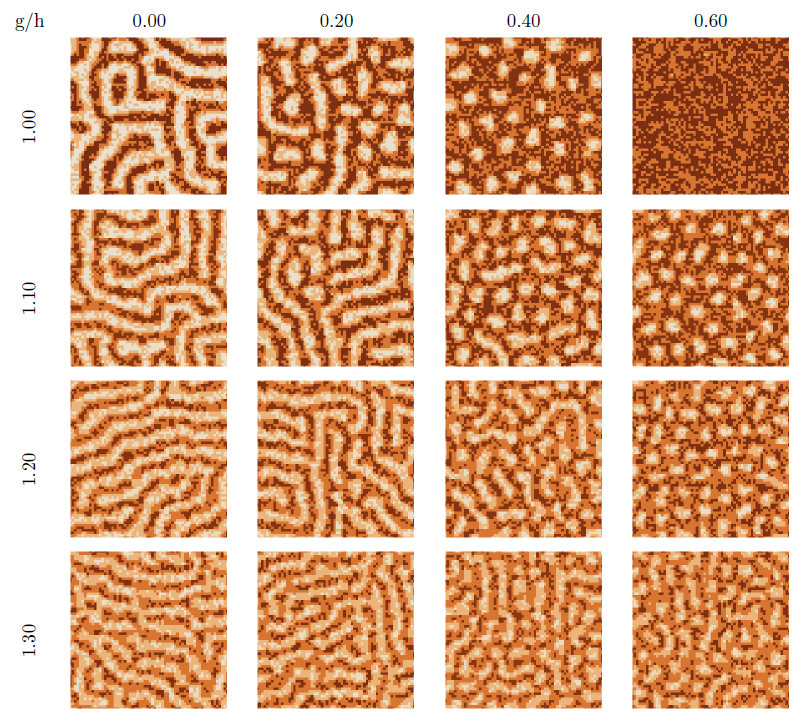}
\caption{Snapshots of equilibrium field configurations $\langle \phi(x) \rangle$ from \eq{simulatable-example} for a range of parameter values $(g,\,h)$ simulated on the lattice. Fifteen of these configurations exhibit patterned behavior. The remaining configuration, in the top left corner, does not exhibit patterns; any variation in color reflects expected small fluctuations. Each patterned configuration exhibits a distribution of shapes. Patterned configurations vary smoothly into one another as we adjust $g$ and $h$, as does the order parameter $\langle \phi_0 \rangle$. As we increase $g$ patterns tend to become thinner and as we increase $h$ the shapes tend to become shorter, until we exit the patterned region. The Fourier transforms of these configurations are shown in \fig{phi4-ft}. The patterned region is in reasonable agreement with the analytic calculations of \sec{3-1}, given the limitations of tree-level analytics, coupled with finite lattice size and spacing.
}\label{fig:phi4-configs}
\end{center}
\end{figure}

In \fig{phi4-configs} we show configurational snapshots of the order parameter $\langle \phi \rangle$ taken at the end of these long runs. The color scale consists of four equally-spaced bins running from approximately -3 to 3, ranging from dark to light. Fifteen of these figures exhibit patterning behavior, and one (top right) does not. This patterned region is in reasonable agreement with the analytic predictions of \sec{3-1}, 
given the limitations of finite lattice size, finite lattice spacing, and tree-level perturbation theory. In particular, the simulation results confirm the overall phase diagram determined analytically.

The patterned field configurations take the form of curved stripes and dots of large positive $\langle \phi \rangle$ value floating in a sea of large negative $\langle \phi \rangle$ value, as in other pattern-forming systems. Typically, there is a small transition region between the two most extreme values in each configuration. A single configuration may contain both dots and stripes of various lengths, with irregular orientation and placement relative to one another.  As we increase $h$ and move rightward through \fig{phi4-configs}, we notice that patterns tend to decrease in characteristic length scale. Twisted and connected stripes gradually shorten into individual strands, which shorten until becoming balls, and eventually these balls too disappear. On the other hand, as we increase $g$ and move downward in the figure, the characteristic width of patterns tends to become smaller until passing the limitations of what we can observe with our lattice size and spacing. As we vary $g$ and $h$, the observed pattern morphologies change smoothly, as does the average action $\langle S \rangle$. 

\begin{figure}
\begin{center}
\includegraphics[width = 6 in]{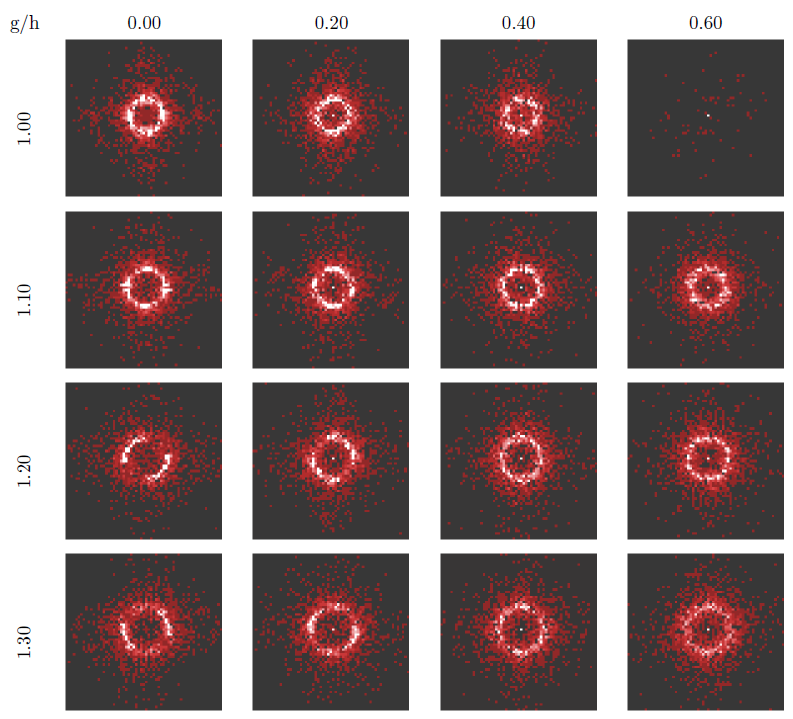}
\caption{Absolute value $|\tilde{\phi}(k)|$ of the Fourier transforms of the position-space field configurations $\langle \phi(x) \rangle$ in \fig{phi4-configs}. Patterned configurations in position space correspond to a ring-shaped configuration in momentum space. The sole non-patterned position-space configuration in \fig{phi4-configs} corresponds to a sharp peak at the origin in momentum space (top right). The presence of rings suggests the common underlying origin of patterning: propagator poles with $q^2>0$. In some cases a small number of modes on the ring are excited, which may occur due to finite size effects, lattice pinning, or locking into an atypical region of configuration space.}\label{fig:phi4-ft}
\end{center}
\end{figure}

In \fig{phi4-ft}, we take the position-space Fourier transform of \fig{phi4-configs} and show its absolute value on a $64\times 64$ momentum lattice with $q=0$ at the center. We clip the maximum value of the Fourier transform at each site at 10 and have colors running from 0 (black) to 10 (white). We immediately notice that the ring-shaped configurations in \fig{phi4-ft} are associated with the patterned configurations in \fig{phi4-configs}. The presence of a momentum-space ring confirms the explanation in \sec{3-1} that $q^2>0$ modes are the common source of pattern formation here. In the sole homogeneous field configuration at the top right, the Fourier transform has a single large value at the origin, representing a nonzero expected value for $\phi$.

\begin{figure}
\begin{center}
\includegraphics[width = 4 in]{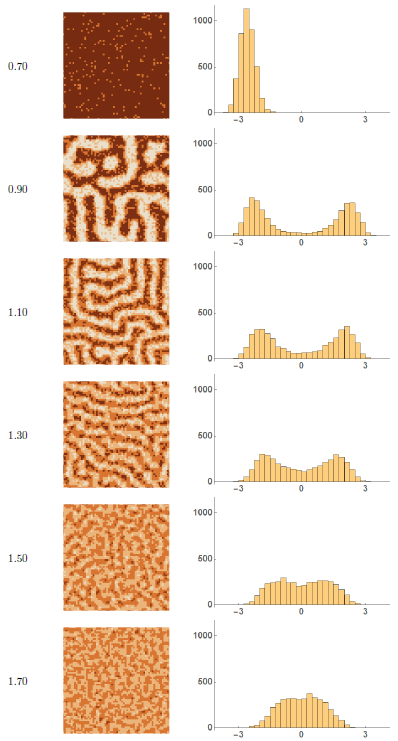}
\caption{A characteristic set of $h=0$ configurations for \eq{simulatable-example} and histograms of the expectation value $|\langle \phi \rangle|$ at each of the $64^2$ values of $x$. The critical line of this system starts at $g=0$ and runs up the $h=0$ axis. The $g=0.70$ configuration exhibits a homogeneous field configuration and asymmetric histogram, and thus lies outside the patterned region. As we move up the $g$ axis to $0.90$, we pass into the patterned region and the histogram takes on a symmetric bimodal form which persists for a large range of $g$ values. Around $g=1.50$ we pass the critical endpoint and into a region where the histogram is a symmetric single-peaked distribution.}\label{fig:phi4-histograms}
\end{center}
\end{figure}

Finally, in \fig{phi4-histograms} we plot a histogram of the expectation value of $\phi(x)$ alongside the configurations which produced the histograms. These configurations all have $h=0$, with $g$ increasing downward. We recall that in the $g$-$h$ plane the critical line lies along $h=0$, running through the normal and complex regions of parameter space before terminating in a critical endpoint inside the patterned region (see \fig{phases2ways}). At $g=0.70$, the field configuration is homogeneous and the histogram shows a single peak centered around a large negative number. This is indicative of the broken symmetry we expect in a non-patterned region. When we move to $g = 0.90$, the field configuration is patterned, the expectation value of the field is zero, and its histogram is bimodal, with the field splitting up into high- and low-valued regions. As we move up the critical line, the two peaks slowly move towards the origin until coalescing into a single hump at the critical endpoint. At $g=1.70$ the curve of the histogram is still fairly broad and we observe patterned field configurations even though the system no longer sits on the critical line. This is consistent with our picture of a critical endpoint in the middle of the patterned region in the $h$-$g$ plane. 

\subsection{Phase transition dynamics} 
Because pattern formation is often seen in phase transition dynamics, it is interesting
to consider the dynamics of our model
as compared with similar behavior seen in phase transition dynamics of
conventional $\phi^4$ models  \cite{cha95,RevModPhys.65.851,doi:10.1080/00018739400101505}.
The order parameter $\phi$ is not conserved, so we can model its dynamics using $S_{\text{eff}}$ and
model A dynamics, {\it i.e.}, with a Langevin equation
\begin{equation}
{\partial \phi( x,t)\over \partial t}=-\Gamma {\delta S_{\text{eff}} \over \delta \phi(x,t)}+\eta(x,t).
\end{equation} 
where $\Gamma$ is a decay constant and $\eta$ is a white noise term.
If the white noise is normalized to $\left<\eta(x,t)\eta(y^\prime,t^\prime)\right>=2\Gamma\delta(x-x^\prime)\delta(t-t^\prime)$,
the distribution of $\phi(x,t)$ will converge for long times to the distribution given by the path integral using $S_{\text{eff}}$.
The difference between our model and the standard model A dynamics of a $\phi^4$ field theory, obtained by taking $g=0$, 
is the nonlocal term in $S_{\text{eff}}$ induced by $\chi$. 
Linearizing the Langevin equation around a homogeneous solution $\phi_0$ and transforming to momentum space, 
the Langevin equation becomes
\begin{equation}
{\partial \tilde\phi( q,t)\over \partial t}=-\Gamma \left(q^2+U''(\phi_0)+{g^2 \over q^2+m_\chi^2}\right)\tilde\phi(q,t)+\tilde\eta(q,t)
=-\Gamma G^{-1}(q)\tilde\phi(q,t)+\tilde\eta(q,t)
\end{equation} 
If the nonlocal term were not present, $\phi_0$ would be unstable when $U''\left(\phi\right)<0$, and spinodal decomposition would occur.
Comparing the dynamics when $g\ne 0$ to the purely local model when $g=0$, we see that the large-$q$ behavior
is identical, but $g\ne 0$ changes the small-$q$ behavior. In the region where pattern formation occurs in equilibrium, 
model A dynamics is that of spinodal decomposition  for large $q$, but relaxational for small $q$. 

We thus interpret the equilibrium patterning behavior of this model as a form of arrested spinodal decomposition.
Starting from a homogeneous solution $\phi_0$ with $U''\left(\phi\right)<0$ , the early-time Langevin evolution will produce the exponentially growing modes
of spinodal decomposition for large $q$, but for small $q$ fluctuations are damped. Spinodal decomposition is arrested at a characteristic scale
in momentum space, with $\phi_0$ stabilized by the nonlocal term.
From this point of view, the chief dynamical difference between the patterned region and the unstable region is that in the patterned region, low $q$
fluctuations are suppressed but grow exponentially with $t$ in the unstable region.

The presence of a characteristic scale, obvious in the Fourier transforms of configuration snapshots,
is reminiscent of model B dynamics for a conserved order parameter \cite{cha95,RevModPhys.65.851,doi:10.1080/00018739400101505}.
Model B dynamics is appropriate for simulations of Ising models when the total magnetization is kept constant, which
is natural when the Ising model is interpreted as a binary allow lattice gas.
The analog for a standard $\phi^4$ models is to hold the average value of $\phi$ fixed.
Linearized model B dynamics of a standard $\phi^4$ model is described by
\begin{equation}
{\partial \tilde\phi( q,t)\over \partial t}=-\Gamma q^2 \left(q^2+U''(\phi_0)\right)\tilde\phi(q,t)+\tilde\eta(q,t).
\end{equation} 
The extra factor of $q^2$ suppresses time evolution of the zero-momentum mode of $\tilde\phi(q,t)$,
which is essentially the average value of $\phi$. In the early stages of spinodal decomposition
with model A dynamics, it is the $q=0$ mode which increases most quickly.
On the other hand, in model
B dynamics the most rapidly growing modes have $q^2>0$, because of the extra factor of $q^2$
in the linearized  Langevin equation. This leads to a characteristic scale in momentum space
for spinodal decompositions with model B dynamics.
In both cases,  the system will
eventually leave the spinodal region $U''\left(\phi\right)<0$ as phase separation completes.
This does not happen in our $\cP\cT$-symmetric model: when $g\ne 0$, large-scale (small-$q$) fluctuations relax quickly,
but large homogeneous regions remain unstable to
fluctuations of nonzero $q$.
In particular, coarsening at arbitrarily large scales does not occur.
\clearpage

\section{$\cPT$ symmetry and QCD at nonzero density}\label{sec:4}

Quantum Chromodynamics (QCD) at nonzero temperature $T$ and chemical potential $\mu$ 
is an important example of a quantum field theory
with a generalized $\cPT$ symmetry. 
Much of what we know about the phase structure of QCD and other gauge theories comes from Euclidean lattice simulations 
combined with a few key theoretical ideas.
The symmetries and order parameters associated with quark confinement and chiral symmetry breaking
are of particular interest as principal determinants of gauge theory
phase structure. None of this changes when the chemical potential is nonzero,
but there are additional complications. The presence of non-positive weights
in the functional integral has proved to be a very difficult problem. We lack
effective algorithms for lattice simulation in this case, and analytical methods must also be modified.
The $\cC\cK$ symmetry of finite-density QCD is a powerful tool.
The sign problem is closely associated with the Euclidean formalism, and
we will focus in this section on quark confinement and its relation to $\cPT$ symmetry.

\subsection{The sign problem and the origin of $\cC\cK$ symmetry in finite density QCD}\label{sec:4-1}


The Euclidean Lagrangian for an SU($N$) gauge field with fermions in the fundamental representation is
\begin{equation}
\mathcal{L}_{\text{QCD}} = \bar{\psi}(i\slashed{D}-m)\psi + \frac{1}{2g^2}\trace(F_{\mu\nu}F^{\mu\nu}).
\end{equation}
The gauge field $A_\mu(x)$ is an SU($N$) Lie algebra-valued field which may be decomposed as $A_\mu(x) = A_\mu^a(x)T^a$, where the $T^a$'s are generators of the group and the index $a$ runs from $1$ to $N^2-1$. The physical case of QCD is three colors of quarks, or $N=3$. The fermion field $\psi$ in general carries flavor. In the case of QCD, the lightest quark flavors are the $u$ and $d$. The gluon field strength tensor $F_{\mu\nu}$ and the covariant derivative $D$ are matrices with color indices, given by
\begin{align}
F_{\mu\nu} &= \left(\partial_\mu A_\nu^a - \partial_\mu A^a_\nu + f^{abc}A_\mu^b A_\nu^c \right)T^a\\
D_\nu &= \partial_\nu + iA_\nu + \mu \delta_{\nu, t}.
\end{align}
The coefficients $f^{abc}$ are the structure constants of SU($N$), and the parameter $\mu$ in the covariant derivative is the chemical potential. Note that the introduction of a nonzero chemical potential not only breaks Euclidean spacetime invariance by establishing a preferred frame,
but leads to the non-Hermitian behavior at finite density: the operator $iD_\nu$ is Hermitian when $\mu=0$, but not when $\mu\ne 0$.

The free energy $F(T,\mu)$ of a gauge theory may be calculated from the partition function $Z$ via $F = -T\log Z$, with $Z$ given by the Euclidean path integral
\begin{align}
Z &= \int [dA_\mu]\, [d\psi] [d\bar{\psi}]\, e^{-S}\nonumber\\
&= \int [dA_\mu]\, \det M(\mu)\,e^{S_{YM}},
\end{align}
where the Fadeev-Popov determinant necessary in continuum perturbation theory is included in the measure $[dA_\mu]$.
In the second line, we have performed the functional integral over the fermion fields, giving rise to the fermion functional determinant $\det M(\mu)$, with $M=i\slashed{D}-m$. We impose nonzero temperature by giving Euclidean spacetime the topology $R^3 \times S^1$, with the circumference of $S^1$ given by $\beta = T^{-1}$. The fermion determinant has an expansion in terms of closed fermion paths through spacetime. Each closed path $p$ carries with it a non-Abelian version of the Aharanov-Bohm phase
\begin{equation}\label{eq:wilson-line}
W(p) = \trace \left\{\cP \exp \left[i\oint_p dx^\mu A_\mu(x) \right] \right\},
\end{equation}
the Wilson loop. Here, $\cP$ indicates path ordering along $p$ of $A_\mu(x)$. For $N\geq 3$ these loops are complex. In the absence of a chemical potential, complex contributions to the fermion determinant from a given Wilson loop are canceled becaue they occur in pairs, e.g. $W(p) + W^*(p)$, tracing the same path but in opposite directions. At nonzero temperatures, Wilson loops may have a nonzero winding number $n$ around the timelike, $S^1$ direction. A winding number $n>0$ represents a net current of fermions in the timelike direction, while $n<0$ represents a net current of antifermions. If $\mu \neq 0$, each Wilson loop in the expansion of the fermion determinant picks up an additional factor of $e^{\beta n \mu}$ so that $W(p) + W^*(p)$ becomes $e^{n\beta \mu} \, W(p) + e^{-n\beta\mu}W^*(p)$, and the imaginary parts do not cancel. Because of this, the fermion determinant gives rise to complex weights in the path integral. These complex weights give rise to the sign problem in finite-density QCD.
This sign problem appears to be an artifact of the Euclidean formalism, but efforts to study finite density QCD using the Hamiltonian formalism
have not provided a compelling alternative. The key role of paths which wind around the $S^1$ direction in Euclidean space indicates
the importance of topology within the Euclidean formalism.

Our understanding of the $\cC\cK$ symmetry of SU($N$) gauge theories at finite-density is closely related to the sign problem. 
Within the Euclidean formalism,  it is most natural to use complex conjugation $\cK$ rather than time reversal $\mathcal T$
as the fundamental antilinear operation. Systems at finite density, with chemical potential $\mu$ nonzero,
explicitly break invariance under the linear charge conjugation operation $\cC$.
Finite density QCD is invariant under the combined action of $\cC$ and $\cK$; this $\cC\cK$ symmetry is the generalized $\cP\cT$ symmetry
of finite density QCD.
Charge conjugation interchanges $W(p)$ and $W^*(p)$, as does complex conjugation $\cK$; in addition, $\cK$ conjugates any complex coefficients in the expansion. Because the fermion determinant is real when $\mu = 0$, we know that the coefficients in the expansion are real. Thus, the combined action $\cC\cK$ leaves the fermion determinant invariant, and finite-density gauge theories are $\cC\cK$ symmetric. If we allow the chemical potential to be complex, an extension of this argument allows us to understand the relation
\begin{equation}
\det M(\mu) = \det M^*(-\mu),
\end{equation}
as a consequence of $\cC\cK$ symmetry, although this identity also may be proven using gamma-matrix identities.

\subsection{Quark confinement and center symmetry}\label{sec:4-2}
The most striking feature of QCD is quark confinement: at low temperature
and density, quarks and gluons are confined inside baryons and mesons.
In this section and the next, we review some well-known basic features of confinement; see, {\it e.g.}, Ref. \cite{Ogilvie:2012is}. for more details.
In lattice simulation of gauge fields without other particles, the Wilson loop
has an area law behavior
\begin{equation}
\left<\trace_{\text{F}}W(p)\right>\sim\exp\left[-\sigma A\right],
\end{equation}
where $\sigma$ is the string tension, which measures
the confining force.
At nonzero temperature, the string tension
may also be measured using the Polyakov loop $P$, also known as the
Wilson line. The Polyakov loop is essentially a Wilson loop that uses
a compact direction in space-time to close the curve using a topologically
nontrivial path in space time.
The Polyakov loop is given by
\begin{equation}\label{eq:polyakov-loop}
P(\vec{x})=\cP\exp\left[i\int_{0}^{\beta}dx_{4}\,A_{4}(x)\right].
\end{equation}
The one-point function of the Polyakov loop measures the free energy $F_Q$
needed to add a heavy quark to the system
\begin{equation}\label{eq:polyakov-one-pt}
\langle \trace_{\text{F}}P(\vec{x})\rangle \sim e^{-\beta F_Q}.
\end{equation}
In a confined phase, $\langle \trace_{\text{F}}\,P(\vec{x})\rangle=0$, which we interpret as $F_Q=\infty$.

Quark confinement is associated with the breaking of a global symmetry
of the gauge action, center symmetry.
The center of a Lie group is the set
of all elements that commute with every other element; the center of SU($N$) is Z($N$).
The action of the gauge field is invariant under global center
symmetry representations because the gauge fields are in the adjoint
representation of SU($N$), on which center symmetry transformations are trivial.
This is particularly clear in lattice gauge theories, where a link
variable $U_{\mu}(x)$ in the gauge group is associated with each lattice site $x$ and
direction $\mu$. The link variable is identified with continuum path-ordered
exponential of the gauge field from $x$ to $x+\hat{\mu}$: $U_{\mu}(x)=\exp\left[iaA_{\mu}(x)\right].$
Consider a center symmetry transformation on all the links in the timelike
direction at a fixed time. In SU($N$) gauge theories at finite temperature
this is $U_{4}(\vec{x},t)\rightarrow zU_{4}(\vec{x},t)$
for all $\vec{x}$ and fixed $t$, with $z\in$ Z($N$). Lattice
gauge actions such as the Wilson action are composed of sums of small Wilson
loops called plaquettes, which are invariant under this global symmetry. However, the
Polyakov loop transforms as $P(\vec{x})\rightarrow zP(\vec{x})$.
Unbroken global Z($N$) symmetry thus implies $\langle \trace_{F}P(\vec{x})\rangle =0$.
Global Z($N$) symmetry defines the confining phase of a pure gauge
theory, and the Polyakov loop is the order parameter for the deconfinement transition
at nonzero temperature.
Above a critical temperature,  $\langle \trace_{\text{F}}\,P(\vec{x})\rangle \ne 0$,
indicating a spontaneous breaking of Z($N$) symmetry.

Fermions in the fundamental representation, such as quarks, explicitly break the Z($N$) symmetry
of a confining gauge system, 
and $F_Q$ is always finite in this case.
If $\mu=0$, charge conjugation symmetry implies that $F_Q=F_{\bar Q}$.
At nonzero density, charge conjugation is explicitly broken, and in general
we expect $F_Q \ne F_{\bar Q}$. This is also the statement
$\langle \trace_{\text{F}}\,P(\vec{x})\rangle \ne \langle \trace_{\text{F}}\,P^\dagger (\vec{x})\rangle$
where both are real.
In a Hermitian theory it would be surprising that the expectation values of two
Hermitian conjugate operators have two different real values: we would expect
that the expectation values would be complex conjugates, perhaps with vanishing
imaginary part. This behavior is readily explained in terms of $\cPT$ symmetry:
the operator $P-P^\dagger$ is odd under $\cC$ and $\cK$ and has a real expectation value.


\subsection{Deconfinement  and Z($N$) symmetry}\label{sec:4-3}

At very high temperatures, perturbation theory is reliable because of asymptotic freedom:
the running gauge coupling becomes arbitrarily small as $T\rightarrow \infty$.
Perturbation theory indicates that Z($N$) symmetry is broken in this region,
even when only gauge bosons are considered.
The one-loop effective
potential for a gauge boson in the background of a static Polyakov
loop $P$ can be easily evaluated in a gauge where $A_{4}$ is time-independent
and diagonal \cite{Gross:1980br,Weiss:1980rj}. 
The contribution added when $T>0$ can be written as
\begin{equation}
\Veff^{1l}\left(P\right)=-\frac{2}{\pi^{2}}\sum_{n=1}^{\infty}\frac{1}{n^{4}}\trace_{\text{Adj}}P^{n}.
\label{eq:V_pure_glue}
\end{equation}
where the factor of two represents the two helicity states of each mode.
This can be interpreted as a sum of contributions from gauge boson worldlines wrapping around the
compact direction an arbitrary number of times.

From this form, it is easy to see that $\Veff^{1l}\left(P\right)$
is minimized when all the moments $\trace_{\text{Adj}}P^{n}$ are maximized. This
occurs when $P\in$ Z($N$), which gives $\trace_{\text{Adj}}P^{n}=N^{2}-1$. This
indicates that the one-loop gluon effective potential favors the deconfined
phase. The pressure $p$ is the negative of the free energy density
at the minimum, 
\begin{equation}
p(T)=2(N^{2}-1)\frac{\pi^{2}T^{4}}{90},
\end{equation}
 which is exactly $p$ for a blackbody with $2(N^{2}-1)$
degrees of freedom. 
From this we see directly the spontaneous breaking of Z($N$) symmetry by gauge bosons
at high temperature.

In many systems, broken symmetry phases occur at low temperatures
and symmetry is restored at high temperatures. The phase structure
of gauge theories as a function of temperature is unusual because
the broken-symmetry phase is the high-temperature phase. A lattice
construction of the effective action for Polyakov loops, valid for
strong-coupling, is instructive \cite{Polonyi:1982wz,Ogilvie:1983ss,Green:1983sd,Gross:1984wb}.
The spatial link variables may be integrated out exactly if spatial
plaquette interactions are neglected. Each spatial link variable then
appears only in two adjacent temporal plaquettes, and may be integrated
out exactly using the same techniques that are used in the Migdal-Kadanoff
real-space renormalization group \cite{Ogilvie:1983ss,Billo:1996pu}.
The resulting effective action has the form
\begin{equation}
S_{\text{eff}}=-\sum_{\left\langle jk\right\rangle }K\,\left[\trace_{\text{F}}P_{j}\trace_{\text{F}}P_{k}^{\dagger}+\trace_{\text{F}}P_{k}\trace_{\text{F}}P_{j}^{\dagger}\right]
\end{equation}
where $K$ is a function of the lattice gauge coupling $g^{2}$ and
the temperature in lattice units: $n_{t}=1/Ta$.
A strong-coupling calculation gives 
a explicit form for $K$ as
$K\simeq\left(1/g^{2}N\right)^{n_{t}}$ to leading order. In the weak-coupling
limit, a Migdal-Kadanoff bond-moving argument gives $K\simeq2N/g^{2}n_{t}$.
This effective action represents a Z($N$)-invariant nearest-neighbor
interaction of a spin system where the Polyakov loops are the spins.
It depends only on gauge-invariant quantities. Standard expansion
techniques show that the Z($N$) symmetry is unbroken for small $K$,
and broken for large $K$. This model explains why the high-temperature
phase of gauge theories is the symmetry-breaking phase: the relation
between $K$ and the underlying gauge theory parameters is such that
$K$ is small at low temperatures and large at high temperatures,
exactly the reverse of a classical spin system where the coupling
is proportional to $T^{-1}$.

The lattice construction of the Polyakov loop effective action is
a concrete realization of Svetitsky-Yaffe universality \cite{Svetitsky:1982gs},
which states that a second-order deconfinement transition in a $(d+1)$-dimensional
gauge theory is in the universality class of classical spin systems
in $d$ dimensions with the same global symmetry. Lattice simulations
indicate that all pure SU($N$) gauge theories have a deconfining
phase transition at some temperature $T_{d}$, above which center
symmetry is broken. In accordance with predictions based on universality,
the deconfinement transition for an SU(2) gauge theory in $3+1$
dimensions has been well established as being in the universality
class of the three-dimensional Ising model, exhibiting a second-order
transition at $T_{d}$. The deconfinement transition for SU(3)
in $3+1$ dimensions is first order. This is consistent with Landau-Ginsburg
predictions for a system with a Z(3) symmetry. The transitions
for $N>3$ appear to be first-order in $3+1$ dimensions as well, with a smooth limit as
$N$ goes to infinity \cite{Lucini:2002ku,Lucini:2003zr}.

\subsection{Quarks and effective actions}\label{sec:4-4}

The addition of quarks to QCD directly affects deconfinement because
of the loss of center symmetry. The effects
can be seen directly in the effective potential for quarks moving in a
nontrivial Polyakov loop background.
The one-loop contribution from nonzero temperature and density can be written
as \cite{Meisinger:2001fi}
\begin{equation}
V_{F}\left(\theta\right)=-\frac{2 N_f}{\beta}\trace_{\text{F}}\int\frac{d^{d}k}{\left(2\pi\right)^{d}}  \left[ \ln\left(1+P e^{\beta\mu-\beta\omega_{k}}
\right)+\ln\left(1+P^\dagger e^{-\beta\mu-\beta\omega_{k}}\right)\right],
\end{equation}
where the factor of two accounts for spin degeneracy, $N_f$ denotes the number of flavors,
and $\omega_k$ is the fermion energy.
The only  change from free fermions at nonzero temperature and density are the factors
of $P$ and $P^\dagger$. 

From a field theory point of view, $V_F$ is a contribution to the effective potential from quarks at nonzero temperature and density.
This has a simple physical interpretation. The Polyakov loop is the non-Abelian analog of the Bohm-Aharanov phase factor
of electromagnetism for a particle winding around the Euclidean time direction at finite temperature.
If we were working in Minkowski space, the Boltzmann weight for a classical particle at $\vec r$ with momentum $\vec k$
and charge $Q$ would be
\begin{equation}
e^{\beta\mu-\beta \omega_k -\beta Q A_0(\vec r)}
\end{equation}
where $A_0$ is the static potential. After rotation to Euclidean space this factor becomes
\begin{equation}
e^{\beta\mu-\beta \omega_k +i \beta Q A_4(\vec r)}.
\end{equation}
This coupling of field to particle gives rise to the familiar Coulomb
interaction between particles at finite temperature.
The non-Abelian analog of this Boltzmann factor for particles is
\begin{equation}
P e^{\beta\mu-\beta\omega_{k}}
\end{equation}
while for antiparticles it is
\begin{equation}
P^\dagger e^{-\beta\mu-\beta\omega_{k}}.
\end{equation}
Thus $V_F$ is the free energy density of quarks in a non-Abelian background field,
and is associated with the quark contribution to the the local pressure via $V_F=-\beta p_F$.

More physically, fermion trajectories are weighted by factors of $e^{\beta\mu}$ and $P$ for each time they wind around the lattice
in the Euclidean time direction, while antifermion trajectories are weighted by factors $e^{-\beta\mu}$ and $P^\dagger$.
In the low-temperature and low-density region, the Polyakov loop factors suppress the contributions of colored excitations.
For $\mu>0$, quark contributions are increased, eventually overcoming the effects of the Polyakov loops
and leading to a change from a dense gas of hadrons to a dense gas of quarks and gluons.
The complex nature of $V_F$ is manifest in the differences between the terms representing
fermions and antifermions: at nonzero density $P$ appears with a factor $e^{\beta\mu}$, while $P^\dagger$ has a factor 
$e^{-\beta\mu}$. 
While complex, $V_F$ is $\cC\cK$-symmetric because both $\cC$ and $\cK$ interchange $P$ and $P^\dagger$.

A variety of models have been used to study the properties of QCD related
to $\cPT$ symmetry.
There is strong evidence from various calculations that sinusoidally-modulated correlation functions,
 a hallmark of $\cPT$-symmetric field theories, appear in models of finite-density QCD.
On the basis of a flux tube model equivalent to a Z(3) spin model with a sign problem, Patel suggested
that there might be an oscillatory signal in two-point baryon number density correlation functions \cite{Patel:2012vn,Patel:2011dp}.
Signals of oscillatory behavior were subsequently found in Polyakov loop correlation functions
using a variety of methods. In \cite{Nishimura:2015lit}, strong-coupling expansions were used for heavy quarks in SU(3) lattice
gauge theory to demonstrate the existence of parameter regions where Polyakov loops have sinusoidally-modulated
exponential decay. Akerland and de Forcrand  demonstrated similar behavior in a Z(3) model with a complex action,
using both mean field theory and lattice simulation \cite{Akerlund:2016myr}.

In \cite{Nishimura:2014rxa,Nishimura:2014kla}, a class of phenomenological continuum models were studied
that combine one-loop thermodynamics with the effects of confinement for the case of SU(3)
gauge bosons and two flavors of quarks at finite temperature and density.
These models are all described by an effective potential which is the sum
of three terms: 
\begin{equation}
\Veff(P)=V_{g}(P)+V_{f}(P)+V_{d}(P).
\end{equation}
The potential term $V_{g}(P)$ is the one-loop effective potential
for gluons, and the potential term $V_{f}(P)$
contains all quark effects, including the one-loop expression
for quark thermodynamics. The potential term $V_{d}(P)$
represents confinement effects. Three different forms
for $V_{f}(P)$ were considered: heavy quarks, massless quarks,
and a Polyakov-Nambu-Jona Lasinio (PNJL)  model  \cite{Fukushima:2003fw}. 
These models generalize Nambu-Jona Lasinio models to include
confinement-deconfinement effects as well as chiral symmetry effects.
The effective potentials of such  models will be a function
of both the chiral order parameter $\bar{\psi}\psi$ and $P$. 
Two different forms for the confining potential $V_{d}(P)$
were used as well, so a total of six different models were studied.
The PNJL models were the most sophisticated models studied, because
they include the effects of chiral symmetry breaking as well as the
deconfinement transition.
In all cases, regions of parameter space at low $T$ and intemediate values of $\mu$
were found where complex-conjugate eigenvalues appear in the Polyakov loop mass matrix.
This is an indicator for $\cPT$ symmetry breaking, and leads to sinusoidally-modulated exponential decay of Polyakov loop correlation functions.
Such regions are separated from the region where normal exponential decay
occurs by disorder lines, where the eigenvalues of the mass matrix move continuously
from real to complex values.
For the PNJL models, the region where complex eigenvalues occurs is associated with
the critical endpoint of a line of first-order phase transitions, similar to the line
in \fig{qcd-phase-diagram}.

Subsequent work \cite{Nishimura:2016yue} showed that the methods used for QCD could be used
to describe the statistical mechanics of a large class of models with liquid-gas transitions using a $\cPT$-symmetric
action. From a field theory perspective, three-dimensional $\mathcal{CK}$-symmetric models are obtained
from dimensional reduction of a four-dimensional field theory at finite
temperature and density.
The simplest cases of interest are models
with a single type of particles, interacting via a scalar field $\sigma$
and a vector field $A_{\mu}.$ Both $\sigma$ and $A_{\mu}$ will
be taken to have masses. The potential induced by $\sigma$ will be
attractive, while that caused by the static vector potential $A_{4}$
will be repulsive between particles. The particles of the underlying
theory are integrated out, and after dimensional reduction and redefinition
of fields, a Lagrangian of the general form
\begin{equation}
L_{3d}=\frac{1}{2}\left(\nabla\phi_{1}\right)^{2}+\frac{1}{2}m_{1}^{2}\phi_{1}^{2}+\frac{1}{2}\left(\nabla\phi_{2}\right)^{2}+\frac{1}{2}m_{2}^{2}\phi_{2}^{2}-F(\phi_{1},\phi_{2})
\end{equation}
is obtained.
Here $\phi_{1}$ is associated with the attractive force and $\phi_{2}$
with the repulsive force. The field $\phi_{1}$ is naturally a
four-dimensional scalar, but $\phi_{2}$ can be obtained from the fourth
component of a vector interaction. The function $F$ can be interpreted
as $\beta p\left(\phi_{1},\phi_{2}\right)$, where $\beta$ is the
inverse of the tempurature $T$ and $p$ is a local pressure. In particular,
$p\left(\phi_{1},\phi_{2}\right)$ is the local pressure of the gas
of particles in the grand canonical ensemble in the presence of the
background fields $\phi_{1}$ and $\phi_{2}$. This is the simplest
class of three-dimensional field theories with liquid-gas phase transitions
that may be derived from renormalizable four-dimensional field theories.
For a classical fluid, both attractive and a repulsive potential terms
are required, as shown by the early example of the van der Waals equation.
In relativistic theories, attractive potentials are associated wth
scalar exchange, but repulsive potential effects require in addition
vector exchange. The use of a massive vector field is natural at nonzero
temperature and density, resulting in a potential between particles
that is the difference of two Yukawa potentials.

The key feature of $L_{3d}$ is that it is not real, but instead satisfies
the $\mathcal{PT}$ symmetry condition 
\begin{equation}
L_{3d}(\phi_{1},\phi_{2})=L_{3d}^*(\phi_{1},-\phi_{2}).
\end{equation}
The charge conjugation transformation $\mathcal{C}$ is the linear
operator which takes charged particles into their antiparticles. This
naturally takes $A_{\mu}\rightarrow-A_{\mu}$ as in the case of QED,
which implies $\phi_{2}\rightarrow-\phi_{2}$; the $\phi_{1}$ field
is left invariant. The antilinear operator used is complex
conjugation. A nonzero chemical potential $\mu$ explicitly breaks
$\mathcal{C}$ symmetry and leads to a complex $L_{3d}$, but the
antilinear symmetry $\mathcal{CK}$ remains \cite{Meisinger:2012va,Nishimura:2014rxa,Nishimura:2014kla,Nishimura:2015lit}.
Because $L_{3d}$ is complex, this class of models has a sign problem.
$\mathcal{PT}$ symmetry implies that the saddle points of $L_{3d}$
have $\phi_{2}$ purely imaginary; at these saddle points, $L_{3d}$
is real. Analytic continuation of the fields into the complex plane
leads to a resolution of the sign problem at tree level, in the sense
that the action is real at the saddle points. More generally, unbroken
$\mathcal{PT}$ symmetry implies that the expected value of $\left\langle \phi_{2}\right\rangle $
must be zero or purely imaginary because $\left\langle i\phi_{2}\right\rangle ^{*}=\left\langle i\phi_{2}\right\rangle $. 
Different models are obtained with different forms for  $F\left(\phi_{1},\phi_{2}\right)$.
For relativistic fermions of
mass $m$, $F$ is taken to be
\begin{equation}
F=\int\frac{d^{3}k}{\left(2\pi\right)^{3}}\log\left[1+\exp\left(-\beta\sqrt{k^{2}+\left(m-g\beta^{-1/2}\phi_{1}\right)^{2}}+\beta\mu+i\beta^{1/2}e\phi_{2}\right)\right].
\end{equation}

\subsection{$\cPT$ symmetry, pattern formation and lattice models of finite-density QCD}

In this section, we apply for the first time the methods of \sec{3} to an effective model of QCD: a static fermion model,
in order to search a pattern-forming region in parameter space.
For lattice gauge theories at finite temperature, the static approximation
represents very well the effect of heavy fermions. In the case where heavy fermions
satisfy $\beta M\gg1$,
their effects may be represented in the action as
\begin{equation}
S_F = - \sum_x 2N_f \log \left\{\det\left[1+ e^{\beta\mu-\beta M}P(x)\right]\det\left[1+ e^{-\beta\mu-\beta M}P^\dagger (x)\right]\right\},
\end{equation}
where we take the sum over space, not spacetime. Note that for $\mu \ne 0$, $S_F$
is complex but $\cC\cK$ symmetric.
This action is often approximated by
\begin{equation}
S_{F}\approx -h_{F}\sum_x \left[e^{\beta\mu}\trace_{\text{F}}P(x)+e^{-\beta\mu}\trace_{\text{F}}P^{\dagger}(x)\right]
\end{equation}
where $h_F = N_f e^{-\beta M}$.
Recalling the spin model interpretation, we see that a heavy particle
behaves like an external field $h_{F}$ coupled to the Polyakov loop.

\begin{figure}
\begin{center}
\includegraphics[width = 3 in]{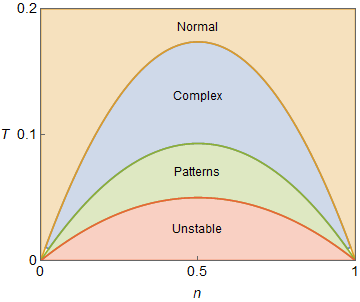}
\caption{Phase diagram of the lattice gas with action \eq{n-T-action} in the $n$-$T$ plane, calculated using the techniques of \sec{3}.
Note the similarity of this phase diagram to \fig{phases2ways}.
}\label{fig:n-T}
\end{center}
\end{figure}

We examine a highly simplified model obtained from lattice models and determine its phase structure. In the fermion determinant,
we make a mean field approximation of the matrix $P(x)$ by $(\trace_{\text{F}} P(x)/N_c)I$, where $N_c$ is the number of colors. Because 
$\trace_{\text{F}} P(x)$ is complex for $N_c \ge 3$, we write $\trace_{\text{F}} P(x)/N_c$ as  $e^{\chi(x) + i\phi(x)}$ and $\trace_{\text{F}}P^\dagger (x)/N_c  = e^{\chi(x) - i\phi(x)}$. We make the further simplifying assumption that antifermion contributions are heavily suppressed by a factor of $e^{-\beta(\mu +M)}$ and ignore them.
As in the liquid-gas models discussed in the previous section, we assume that the field $\chi$ gives rise to an attractive interaction $V_A$ between static fermions, while the field $\phi$ induces a repulsive interaction $V_R$. As before, the simplest choice for $V_A$ and $V_R$ is two Yukawa potentials, and the complete lattice action is
\begin{equation}\label{eq:n-T-action}
S=\frac{1}{2\beta}\sum_{x,y} \left\{\frac{1}{g_\chi^2}\left[(\nabla\chi)^2 + m_\chi^2\chi^2 \right]+\frac{1}{g_\phi^2}[(\nabla\phi)^2 + m_\phi^2\phi^2]\right\}-2N_c N_f \sum_x \log\left[1+z e^{\chi(x)+i\phi(x)}\right],
\end{equation}
where $z = e^{\beta(\mu -M)}$.
This model has a clear relation to the sine-Gordon model. This is perhaps unsurprising because the sine-Gordon model is itself equivalent to a classical Coulomb gas. It was first used to describe a lattice gas with attractive and repulsive interactions by Fisher and Park \cite{PhysRevE.60.6323}, who showed that this model can describe a standard liquid gas transition for $z>0$. However, they were interested in the critical singularity associated with hard-core repulsion which occurs when $z<0$ and showed explicitly that this critical behavior is in the $i\phi^3$ universality class.
We are interested in the normal liquid-gas transition for $z>0$, which is in the Z(2) universality class, and possible patterned phases. We can apply the methods of \sec{3} directly to find the behavior of \eq{n-T-action}. We set $2N_f N_c=1$ to match the work of Fisher and Park and take the parameter values $m_\chi/m_\phi = 0.2$, $g_\chi^2/m_\chi^2 = 0.3$, and $g_\phi^2/m_\phi^2 = 0.5$. The phase diagram in $n$-$T$ space is shown in \fig{n-T}. We note that the phase diagram of \eq{n-T-action} looks much like that of our $\cPT$-symmetric extension of the $\phi^4$ model, \eq{action}, consisting of all four expected regions of behavior: normal, unstable, complex, and patterned. We take this as an indication of the ubiquity of pattern formation in $\cPT$-symmetric quantum field theories and the possibility of such patterns in finite-density QCD.

\section{Concluding remarks and future directions}\label{sec:5}

The inherent non-Hermiticity of finite-density QCD has caused major obstacles to understanding its phase structure,
manifested by a complex action and a sign problem.
The invariance of finite-density QCD under the combined operations of $\cC\cK$
is a powerful tool.  We have developed techniques to determine the phase diagram of scalar $\cPT$-symmetric
quantum field theories.
These techniques may be applied to effective models of the deconfinement transition and hold promise for deepening our understanding of QCD phase structure.

In \sec{2}, we reviewed a recently developed technique to cast a large class of $\cPT$-symmetric quantum field theories  with complex
actions into real positive forms that may be simulated on the lattice. We can now construct arbitrarily many new $\cPT$-symmetric quantum field theories both as extensions of Hermitian universality classes, as well as an indirect construction from the real form. This method may also shed new light on the relationship between $\cPT$-symmetric models, higher derivative theories, and other real models with local and nonlocal interactions. 

In \sec{3}, we examined the phase diagram of $\cPT$-symmetric quantum field theories. We described the standard procedure for finding the stable and unstable regions of a conventional quantum field theory and discussed how this technique breaks down in the non-Hermitian case. We derived the generalization of this technique to the $\cPT$-symmetric case, and confirmed its validity 
for a $\cPT$-symmetric extension of $\phi^4$ theory using lattice simulations. $\cPT$-symmetric scalar field theories may exhibit pattern formation in the vicinity of critical points, a feature impossible in conventional field theories.

\Sec{4} described the origin of $\cC\cK$ symmetry in QCD at nonzero chemical potential in terms of the Polyakov loop,
and the strong evidence for $\cPT$-breaking behavior, manifesting as sinusoidal modulation of the Polyakov loop two-point
function. We also exhibited a simplified model of finite-density QCD with a patterning phase.
It is possible that finite-density QCD exhibits patterned fields in its critical region, which manifest as agglomerations of deconfined matter floating in a sea of confined phase, and vice versa. 
These possibilities pose new challenges and opportunities for theory, experiment, and astrophysical observation.

\section*{Acknowledgments}
STS was supported by a Graduate Research Fellowship from the U.S. National Science Foundation under Grant No. 1745302; the U.S. Department of Energy, Office of Science, Office of Nuclear Physics, from DE-SC0011090; and fellowship funding from the MIT Department of Physics. MCO would like to acknowledge helpful discussion with Hiromichi Nishimura.

\section*{References}

\bibliographystyle{iopart-num}
\bibliography{phhqp-semifinal}

\end{document}